\documentclass[12pt]{iopart}

\usepackage{graphicx}

\usepackage{iopams}

\usepackage[dvipsnames,usenames]{color}

\usepackage{soul}

\newcommand{\eq}[1]{Eq.~(\ref{#1})}

\newcommand{\ket}[1]{|#1\rangle}
\newcommand{\bra}[1]{\langle#1|}

\newcommand{\rhoab}{{\hat{\rho}_{AB}}}

\begin{document}

\title[Quantum discord from system-environment correlations]{Quantum discord from system-environment correlations}

\author{R. Tatham and N. Korolkova}
\address{School of Physics and Astronomy,
 University of St. Andrews,  North Haugh,  St. Andrews,  KY16 9SS,  Scotland,  UK}

\ead{$^\ast$ nvk@st-andrews.ac.uk (corresponding author)}
\begin{abstract}
In an initially uncorrelated mixed separable bi-partite system, quantum correlations can emerge under the action of a local measurement or local noise \cite{Streltsov2011}. We analyze this counter-intuitive phenomenon using quantum discord as a quantifier. We then relate changes in quantum discord to system-environment correlations between the system in a mixed state and some purifying environmental mode using the Koashi-Winter inequality. On this basis, we suggest an interpretation of discord as a byproduct of transferring entanglement and correlations around the different subsystems of a global pure state.

\end{abstract}

\pacs{03.67.Hk, 03.67.Dd, 42.50.Nn}
\maketitle

Quantum discord is a measure of correlations in bi-partite systems that aims to capture more general signatures of quantumness in composite systems than entanglement and can be found even in systems in separable mixed states. These signatures are associated, for example, with the non-commutativity of quantum observables and with the fact that local measurements generally induce some disturbance on quantum states. Much work has been done to determine how well these non-classical correlations represent a physical resource that can be used to improve upon classical performance. A great body of work has also been undertaken to better understand the nature of the non-classical correlations and the mechanisms through which these correlations manifest themselves.

In a recent publication by Streltsov \textit{et al} \cite{Streltsov2011}, it has been shown that if two subsystems in mixed quantum states initially possess no quantum correlations, it is still possible to create non-classical correlations between them by performing a local (quantum) operation on one subsystem - a remarkable and counter-intuitive phenomenon. However, a clear insight into this mechanism can be gained if we incorporate the environment into the picture by, for example, exploring how the original bipartite system is coupled to an external quantum object representing the environment. We can then take a closer look at the behaviour of the non-classical correlations in what is now a pure tri-partite global quantum state.

Quantum discord is a measure of the nonclassicality of correlations in a bipartite quantum state $\hat{\rho}_{AB}$
defined as a difference \cite{Olivier_01}
\begin{equation}\label{D1}
\mathcal{D}^\leftarrow(\hat{\rho}_{AB})={\cal I}_{q}(\hat{\rho}_{AB})-\mathcal{J}^\leftarrow(\hat{\rho}_{AB})
\end{equation}
between the quantum generalizations of two classically equivalent
expressions for the mutual information, namely the quantum mutual information
\begin{equation}\label{I} {\cal I}_{q}(\hat{\rho}_{AB})={\cal
S}(\hat{\rho}_{A})+{\cal S}(\hat{\rho}_{B})-{\cal S}(\hat{\rho}_{AB}),
\end{equation}
where ${\cal S}(\hat{\rho})=-\mbox{Tr}(\hat{\rho}\log_2\hat{\rho})$ is the von Neumann
entropy, and the so called one-way classical correlation \cite{Vedral_01}
\begin{eqnarray}\label{J}
{\cal J}^\leftarrow(\hat{\rho}_{AB})={\cal S}(\hat{\rho}_{A}) - \inf_{\{\Pi_i\}} {\cal
H}_{\{\Pi_i\}}(A|B),
\end{eqnarray}
where ${\cal H}_{\{\Pi_i\}}(A|B){\equiv}\sum_{i}p_{B}(i){\cal
S}(\hat{\rho}_{A|i})$ is the conditional entropy of subsystem $A$ given that a positive operator valued measurement (POVM)
$\{\Pi_{B}(i)\}$ has been performed on subsystem $B$. 
Here $\hat{\rho}_{A,B}=\mbox{Tr}_{B,A}[\hat{\rho}_{AB}]$ are reduced states of
subsystems $A$ and $B$ respectively, $\hat{\rho}_{A|i} =
\mbox{Tr}_B[\Pi_{B}(i)\hat{\rho}_{AB}]/p_{B}(i)$ is the conditional state obtained
upon detecting the POVM element $\Pi_{B}(i)$ on $B$ and
$p_{B}(i)=\mbox{Tr}[\Pi_{B}(i)\hat{\rho}_{AB}]$ is the corresponding
probability. The arrow in the notation indicates on which subsystem the POVM is performed. Subtracting \eq{J} from \eq{I} one gets for the quantum
discord the following formula:
\begin{equation}\label{D2}
\mathcal{D}^\leftarrow(\hat{\rho}_{AB})={\cal S}(\hat{\rho}_{B})-{\cal
S}(\hat{\rho}_{AB})+\inf_{\{\Pi_i\}}{\cal
H}_{\{\Pi_i\}}(A|B).
\end{equation}
Note that quantum discord is generally asymmetric under an exchange of the subsystems.

Streltsov \textit{et al} \cite{Streltsov2011} considered a maximally mixed  two qubit state which for ease of notation we will denote $\hat{\rho}_{AC}$ and assume that the two marginal density matrices $\hat{\rho}_{A,C}=\mbox{Tr}_{C,A}[\hat{\rho}_{AC}]$ are held by Alice and Charlie. The initial state is described by
\begin{equation}
\label{ACin}
\hat{\rho}_{AC}=\frac{1}{2}\left(\ket{00}\bra{00}+\ket{11}\bra{11}\right),
\end{equation}
which is clearly separable and exhibits no quantum correlations. A local measurement is then performed on Charlie's subsystem yielding the state
\begin{equation}
\label{ACout}
\hat{\rho}_{A'C'}=\frac{1}{2}\left(\ket{00}\bra{00}+\ket{1+}\bra{1+}\right)
\end{equation}
where $\ket{+}=(1/\sqrt{2})\left(\ket{0}+\ket{1}\right)$. This local operation has the effect of establishing non-zero quantum discord in one direction whilst leaving the quantum discord as zero in the other direction. In particular, $\mathcal{D}^\leftarrow(\hat{\rho}_{A'C'})>\mathcal{D}^\rightarrow(\hat{\rho}_{A'C'})=0$ after the local operation on subsystem $C$. We would like to consider why this result occurs for this mixed state by considering the state as part of a pure, global tri-partite state.
 
We first note that a mixed state is \textit{not} a fundamental object. A mixed state is simply a sign of ignorance on the behalf of the observer. In the simplest case, an incoming qubit may interact with the system under consideration and then be lost. In the interaction process, correlations both classical and quantum will be formed between the studied system and the environment depending on the type of interaction. In actual fact, by trying to find a purification of the state under consideration, one is simply trying to account for any interactions with the environment. It does not matter if one qubit or more has interacted with the system - we combine all the environment qubits that ever were lost into another system $B$, held by Bob, represented by one qubit but conveying the statistical information of them all, such that $\textrm{Tr}_B\left[\hat{\rho}_{ABC}\right]=\hat{\rho}_{AC}$. In the example above~(\ref{ACin}), one purification is given by the GHZ state \cite{GHZ}:
\begin{equation}
\label{GHZ}
\ket{\Psi}_{ABC}=\frac{1}{\sqrt{2}}\left(\ket{000}+\ket{111}\right)
\end{equation}
which gives the mixed state of Eq~(\ref{ACin}) after tracing out subsystem $B$. As the global state is pure, the entropies are the same 
across any bi-partition, so
\begin{eqnarray}
\label{entequ}
\mathcal{S}\left(\hat{\rho}_A\right)=\mathcal{S}\left(\hat{\rho}_{BC}\right),\quad\mathcal{S}\left(\hat{\rho}_B\right) = \mathcal{S}\left(\hat{\rho}_{AC}\right),  \quad \mathcal{S}\left(\hat{\rho}_C\right) = \mathcal{S}\left(\hat{\rho}_{AB}\right).
\end{eqnarray}
Initially, there is entanglement across each of the bipartitions of the GHZ state as quantified by the Entanglement of Formation $\mathcal{E}_F$  \cite{Wootters1998}:
\begin{equation}
\mathcal{E}_F\left(\hat{\rho}_{AB,C}\right)=\mathcal{E}_F\left(\hat{\rho}_{A,BC}\right)=\mathcal{E}_F\left(\hat{\rho}_{B,AC}\right)=1.
\end{equation}
If knowledge of any one of the three subsystems is missing (any of the subsystems is traced out), there is no entanglement between the two remaining subsystems. However, {\it classical} correlations between those that remain are then maximal as quantified by the one way classical correlations $\mathcal{J}^\leftarrow$. Consequently,
\begin{equation}
\mathcal{E}_F\left(\hat{\rho}_{AB}\right)=\mathcal{E}_F\left(\hat{\rho}_{AC}\right)=\mathcal{E}_F\left(\hat{\rho}_{BC}\right)=0;
\end{equation}
\begin{equation}
\mathcal{J}^\leftarrow\left(\hat{\rho}_{AC}\right)=\mathcal{J}^\rightarrow\left(\hat{\rho}_{AC}\right)=1, \label{Cl-corr-AC}
\end{equation}
\begin{equation}
\mathcal{J}^\leftarrow\left(\hat{\rho}_{AB}\right)=\mathcal{J}^\rightarrow\left(\hat{\rho}_{AB}\right)=1,
\end{equation}
\begin{equation}
\mathcal{J}^\leftarrow\left(\hat{\rho}_{BC}\right)=\mathcal{J}^\rightarrow\left(\hat{\rho}_{BC}\right)=1
\end{equation}
and
\begin{equation}
\mathcal{S}\left(\hat{\rho}_{A}\right)=\mathcal{S}\left(\hat{\rho}_B\right)=\mathcal{S}\left(\hat{\rho}_C\right)=1.
\end{equation}
Koashi and Winter \cite{Koashi2004} showed that, in any pure tripartite system, these quantities are intimately related as
\begin{equation}
\label{Koashi}
\mathcal{S}\left(\hat{\rho}_A\right)=\mathcal{E}_F\left(\rhoab\right)+\mathcal{J}^\leftarrow\left(\hat{\rho}_{AC}\right).
\end{equation}
Thus in a pure tripartite system, as a consequence of the above relations, if subsystem $A$ is maximally entangled with subsystem $B$, then subsystem $A$ cannot be even classically correlated with subsystem $C$ (and vice versa). Hence, as Koashi and Winter discussed, the entropy of the marginal $\hat{\rho}_A$ can be thought of as the capacity of Alice's subsystem to form correlations.

Let us now turn to our example and consider what must happen after a local measurement on $C$. If Alice had no knowledge of Bob's or Charlie's state, then the probability distribution yielded from her part of the state $\hat{\rho}_A=\textrm{Tr}_{BC}\left[\hat{\rho}_{ABC}\right]$ will stay unchanged. In other words, she implicitly averages over everything that could happen to states $B$ or $C$ by not knowing anything about them. Consequently, after the measurement on subsystem $C$, $\mathcal{S}\left(\hat{\rho}_{A'}\right)=\mathcal{S}\left(\hat{\rho}_{A}\right)$. A similar argument holds for Bob and so $\mathcal{S}\left(\hat{\rho}_{B'}\right)=\mathcal{S}\left(\hat{\rho}_{B}\right)$.

Note that the purity must be preserved upon the measurement on any of the subsystems, $\textrm{Tr}\left[\hat{\rho}^2_{A'B'C'}\right]=\textrm{Tr}\left[\hat{\rho}^2_{ABC}\right]=1$, provided nothing more is lost to further environment qubits.
The only local distribution to be affected by the local measurement in this case is that of Charlie's subsystem. After the local operation, the entropy of $\hat{\rho}_{C'}$ is given by:
\begin{equation} \label{reducedEntropy}
\mathcal{S}\left(\hat{\rho}_{C'}\right)=\frac{\ln\left[8\right]-\sqrt{2}\coth^{-1}\left[\sqrt{2}\right]}{\ln\left[4\right]} == S^C_0.
\end{equation}
The entropy of Charlie's subsystem has decreased, $\mathcal{S}\left(\hat{\rho}_{C'}\right)<\mathcal{S}\left(\hat{\rho}_{C}\right)$. That is, the capacity of the quantum state that Charlie holds to form correlations has decreased. As the state $\hat{\rho}_C$ was initially separable from $\hat{\rho}_A$, a local operation on $\hat{\rho}_C$ cannot create entanglement between $A$ and $C$. Thus, in this case, the only correlations that $\hat{\rho}_{C'}$ can form are classical, and they are weaker than before the measurement:
\begin{eqnarray} \label{entropy-change}
\mathcal{S}\left(\hat{\rho}_{C'}\right)=\mathcal{J}^\rightarrow\left(\hat{\rho}_{B'C'}\right)= \mathcal{J}^\rightarrow\left(\hat{\rho}_{A'C'}\right)= S^C_0. 
\end{eqnarray}
However, Alice and Bobs' local distributions remain unchanged - they cannot change as they have never been interacted with. Thus their respective capacities for forming correlations remain unchanged: $\mathcal{S}\left(\hat{\rho}_{A'}\right)=\mathcal{S}\left(\hat{\rho}_{A}\right)=1$ and $\mathcal{S}\left(\hat{\rho}_{B'}\right)=\mathcal{S}\left(\hat{\rho}_{B}\right)=1$. According to Eq.~(\ref{entequ}), this leaves $\mathcal{S}\left(\hat{\rho}_{BC}\right)=\mathcal{S}\left(\hat{\rho}_{AC}\right)=1$ unchanged. 

The entanglement between subsystems $A$ and $C$ remains unchanged at zero, although the classical correlations between $A$ and $C$ as measured by $\mathcal{J}^\leftarrow\left(\hat{\rho}_{A'C'}\right)$ and $\mathcal{J}^\rightarrow\left(\hat{\rho}_{A'C'}\right)$ have decreased from unity to $1-S^C_0$ and $S^C_0$ respectively. Feeding this into the Koashi-Winter relation:
\begin{equation}
\mathcal{S}\left(\hat{\rho}_{A'}\right)=\mathcal{E}_F\left(\hat{\rho}_{A'B'}\right)+\mathcal{J}^\leftarrow\left(\hat{\rho}_{A'C'}\right),
\end{equation}
we can see that the remaining capacity for Alice's correlations must be filled up to unity - by Alice's state becoming entangled with Bob's with $\mathcal{E}_F\left(\hat{\rho}_{A'B'}\right)=S^C_0$.

\begin{figure*}
\begin{center}
\includegraphics[width=8cm]{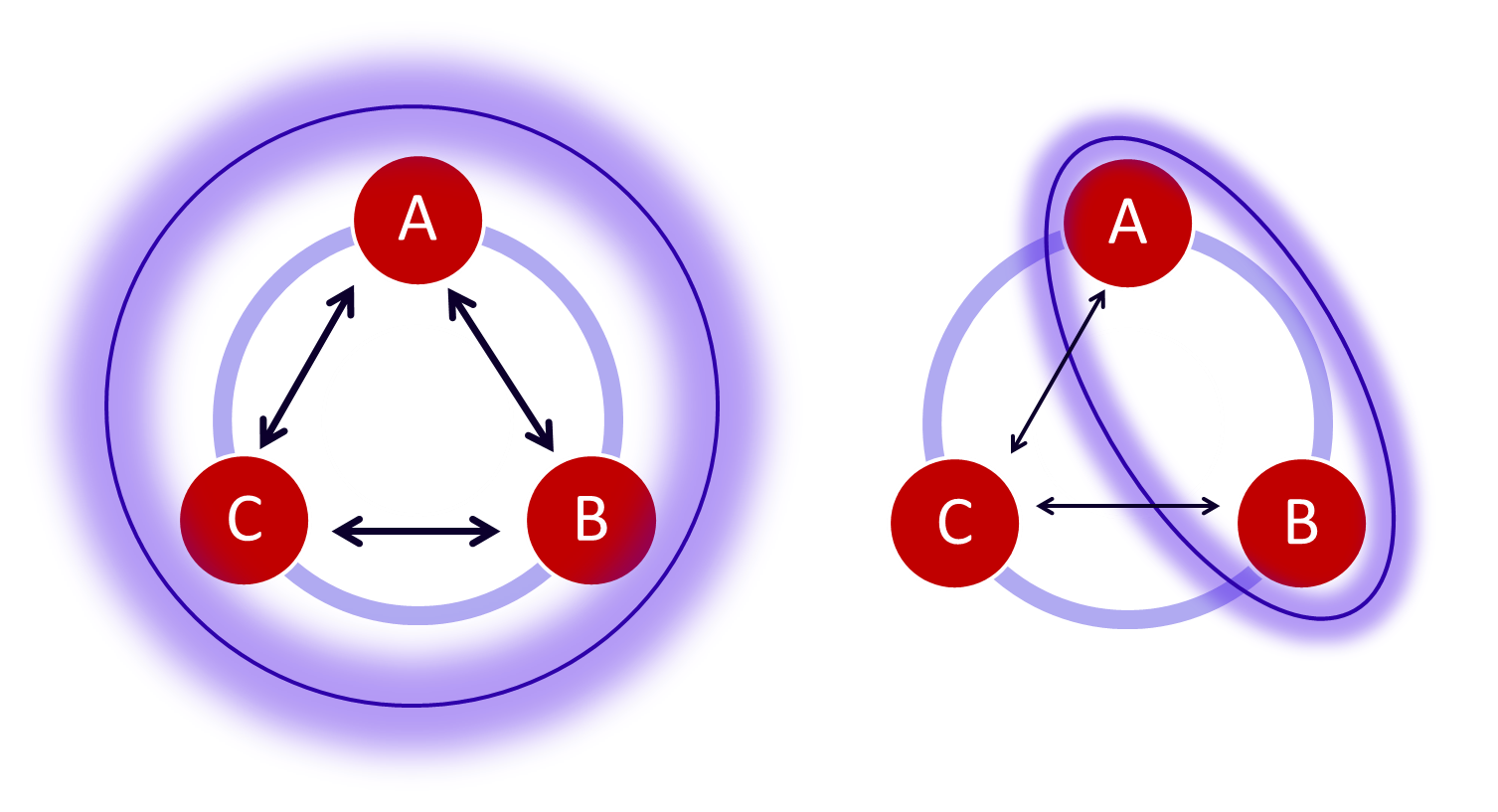}
\caption{\small Initially no two subsystems are entangled (blue glowing circle) but all systems are strongly classically correlated (black arrows) and the entanglement across any bipartition of the pure state is maximum. After the local operation on $C$, the entropy of C decreases. No entanglement can form between $A$($B$) and $C$ as there was none to begin with so only classical correlations can be there, and weaker than previously. However, the capacity of $A$($B$) to be correlated is not used up by this so entanglement must emerge between $A$ and $B$. From the other direction, the local measurement on $C$ can be seen as a non-local measurement on $A$ and $B$.}
\end{center}
\end{figure*}

In other words, a local operation on Charlie's subsystem has preserved the purity of the global state, but decreased the entanglement across the $(AB)-C$
bipartition, $\mathcal{E}_F\left(\hat{\rho}_{A'B',C'}\right)<\mathcal{E}\left(\hat{\rho}_{AB,C}\right)$. The entanglement $\mathcal{E}_F\left(\hat{\rho}_{A,BC}\right)$ across the $A-(BC)$ bipartition and $\mathcal{E}\left(\hat{\rho}_{B,AC}\right)$ across the $B-(AC)$ partition remains the same. This results in reduced classical correlations between Alice (Bob) and Charlie. However, the capacity of Alice (Bob) for forming correlations is unchanged so the consequence is that the subsystems of Alice and Bob become entangled. This is verified by direct calculation using the formula from \cite{Wootters1998}. The discord between $A$ and $C$ arises as a side effect of this entanglement formation between the two subsystems that were not directly altered by the POVM - the mutual information $\mathcal{I}_q\left(\hat \rho_{A'C'}\right)$ is reduced from 1 to $S^C_0$ but one-way classical correlations are decreased further. In fact,
\begin{eqnarray} 
\mathcal{D}^\leftarrow\left(\hat{\rho}_{A'C'}\right)&=&\mathcal{I}_q\left(\hat \rho_{A'C'}\right)-\mathcal{J}^\leftarrow\left(\hat{\rho}_{A'C'}\right)=2S^C_0-1
\nonumber \\
\mathcal{D}^\rightarrow\left(\hat{\rho}_{A'C'}\right)&=&\mathcal{I}_q\left(\hat \rho_{A'C'}\right)-\mathcal{J}^\rightarrow\left(\hat{\rho}_{A'C'}\right)=0. \label{discordLocal}
\end{eqnarray}

Interestingly, we could think about this from the opposite perspective - as opposed to performing a local operation on $C$ we could perform a complementary non-local entangling operation directly on $A$ and $B$. That is, a local operation on $C$ that creates discord between $A$ and $C$ is \textit{equivalent} to an entangling operation on $AB$. This is due to the fact that the global state $\hat{\rho}_{ABC}$ (purification of the initial mixed state $AC$ including environment $B$) is pure before and after the measurement:
\begin{equation}
\mathcal{S}\left(\hat{\rho}_{AB}\right)=\mathcal{S}\left(\hat{\rho}_C\right)
\end{equation}
\begin{equation}
\mathcal{S}\left(\hat{\rho}_{A'B'}\right)=\mathcal{S}\left(\hat{\rho}_{C'}\right).
\end{equation}
The initial transformation was of the form $\hat{\rho}_{A'B'C'}=\hat{E}\hat{\rho}_{ABC}\hat{E}^T$ where
\begin{equation}
\hat{E}=\mathbf{1}_A\otimes\mathbf{1}_B\otimes\left(\begin{array}{cc}1&1/\sqrt{2}\\0&1/\sqrt{2}\\\end{array}\right).
\end{equation}
but could also be performed by a transformation $\hat{E}'\rho_{ABC}\hat{E}^{'T}$ where
\begin{equation}
\hat{E}'=\hat{M}_{AB}\otimes\mathbf{1}_C
\end{equation}
and 
\begin{equation}
M_{AB}=\left(\begin{array}{cccc}1&0&0&0\\0&0&0&0\\0&0&0&0\\1/\sqrt{2}&0&0&1/\sqrt{2}\\\end{array}\right).
\end{equation}
Note, the transformations $\hat{E}$ and $\hat{E}'$ are \textit{not} unitary.

The above considerations unveil quantum discord as being not really a fundamental phenomenon but a side effect of all the changes in local entropy in a quantum system coupled to the environment. Initially, there are only classical correlations across the ($A$-$C$) bi-partition and $\mathcal{D}^\leftarrow\left(\hat{\rho}_{AC}\right)= 0$. If we incorporate the environment into the picture, the global pure tri-partite state is entangled (but there is no entanglement across any of the bi-partitions).  The measurement destroys this entanglement redirecting the entropy flow and resulting in: (1) bi-partite entanglement between  subsystem $A$ and part of enviroment $B$ (purifying subsystem) and (2) bi-partite quantum correlations between the separable subsystem $A$ and the measured subsystem $C$ resulting in non-zero quantum discord $\mathcal{D}^\leftarrow\left(\hat{\rho}_{A'C'}\right)\neq 0$.

This research has been supported by the Scottish Universities Physics Alliance (SUPA) and by the Engineering and Physical Sciences Research Council (EPSRC).

\end{document}